\documentclass[aps,prl,showpacs,twocolumn,superscriptaddress]{revtex4-1}  

\usepackage{mathrsfs,amsmath,amssymb,amsthm,natbib}
\usepackage{amssymb,fmtcount}
\usepackage{graphicx,epsfig,latexsym,overpic,amssymb,color}
\usepackage{threeparttable}
\preprint{\today}

\begin{document}
\title{
Physics Informed Bayesian Machine Learning of Sparse and Imperfect  Nuclear Data
}
\author{Jiaming Liu}
\affiliation{
State Key Laboratory of Nuclear Physics and Technology, School of Physics,
Peking University, Beijing 100871, China
}

\author{Yang Su}
\affiliation{
China Nuclear Data Center, China Institute of Atomic Energy, Beijing 102413, China
}
\author{N.C. Shu}
\affiliation{
China Nuclear Data Center, China Institute of Atomic Energy, Beijing 102413, China
}
\author{Y.J. Chen}
\affiliation{
China Nuclear Data Center, China Institute of Atomic Energy, Beijing 102413, China
}
\author{J.C. Pei}\email{peij@pku.edu.cn}
\affiliation{
State Key Laboratory of Nuclear Physics and Technology, School of Physics,
Peking University, Beijing 100871, China
}
\begin{abstract}
The prevailing data-driven machine learning has been plagued by the absence of physics knowledge and the scarcity of data.
We implement the physics-model informed prior into Bayesian machine learning to evaluate
the energy dependence of independent fission product yields, which are crucial for advanced nuclear energy applications
but only sparse and imperfect experimental data are available. The informative prior is the posterior after learning the generated data from fission models.
Furthermore, cumulative fission yields  are included  as  a physical constraint via a conversion matrix
to provide augmented energy dependence. Our work demonstrated a truly Bayesian machine learning by incorporating comprehensive physics knowledges
as a powerful tool
to exploit the sparse but expensive nuclear data.

\end{abstract}
\maketitle

\emph{Introduction.}---
Over the past decade, data-driven machine learning has emerged as a transformative tool
in scientific research across diverse disciplines. Machine learning has wide applications in
nuclear physics~\cite{rmp,lowenergy}, such as inferences of nuclear observables~\cite{witek,niu,liqf, qiankun,baicl,manana,sunxj,luoya}, solve ill-posed inverse problems~\cite{east,emulator1,inverse2},
emulators of costly theoretical models~\cite{emulator2,emulator3}, and analysis of experimental data~\cite{Neudecker,beam}.
Despite these successes, nuclear physics often confronts small datasets, so that
it wasn't deemed as a good playground for machine learning.
Moreover, nuclear reaction data~\cite{reactiond,gooden} are generally incomplete, noisy and discrepant,
while nuclear structure data~\cite{niu2,liqf} have better accuracies but heavily rely on
non-trivial quantum effects.
To overcome these issues in purely data-driven machine learning, physics-informed, or physics-guided,
or physics-constrained machine learning have been proposed~\cite{pinn}.
Indeed, without embedding physical information, it would be impossible to exploit
 maximum values of scarce data, for instance the gravitational wave observations~\cite{gw170817}.
There are several typical strategies to inform machine learning by using feature inputs including empirical shell and odd-even effects~\cite{niu2,qiao1,qiaonst,schunck2,lovell-mass,liqf,mayg},
or learning the residual error of physics models\cite{witek,wangza1},  or penalized learning with constraints~\cite{wangza3}.
It is known that physics informed neural network (PINN) has been very successful in solving differential equations~\cite{pinn,pinn2}.
However, to inform neural networks with comprehensive physics knowledge is another matter and is particularly needed in nuclear physics.
In principle, Bayesian machine learning has advantages in incorporating
 physics priors~\cite{jpg} and handling imperfect data~\cite{wangza1}.
The objective of this Letter is to inform the priors of Bayesian machine learning by physics models, which contain
comprehensive conservation laws and quantum effects consistently,  for
 evaluations of sparse and imperfect nuclear data.

The supply of high quality nuclear data is crucial for applications of advanced nuclear energy with new types of fuel
and medical isotope productions~\cite{nucleardataneeds}.
Conventionally the recommended nuclear data~\cite{jendl,jeff,endf,cendl} for application needs is from
a combined evaluation of theoretical models and measured data, which highly
depends on experts. In this context, accurate automated and bias-free nuclear data evaluation
are expected by leveraging powerful machine learning approaches.
The evaluation of nuclear fission data is among the most challenging tasks.
Microscopic fission theories such as time-dependent density functional theory (TD-DFT)
is suitable for descriptions of the partition of particles, energies, angular momentum between two nascent fragments,
but not well positioned to reproduce the distributions of fission product yields~\cite{tddft}.
The time-dependent generator-coordinate method (TD-GCM) can reasonably describe
the distributions of fragment mass yields~\cite{tdgcm,tdgcm2} but is not adequate for two-dimensional distributions~\cite{schunck1}.
On the other hand, there have been widely used semi-empirical models with adjustable parameters
for evaluation purposes in specific ranges of nuclear masses and energies~\cite{Wahl,GEF,Brosa}.
The fission product yields vary from primary yields of nascent fragments,
to independent yields after prompt neutron emissions, and to cumulative yields after $\beta$-decays~\cite{fissiontheory}.
Experimental fission yields, particularly from neutron induced fission,  are sparse and have large uncertainties,
mainly due to the lack of monoenergetic neutron sources~\cite{gooden}.
In major evaluation libraries of fission yields, only data with neutron incident energies
at thermal energy (0.0253 eV), 0.5  and 14 MeV are available~\cite{jendl,jeff,endf,cendl}.
Even these evaluated data are acceptable, the interpolation of energy dependence and
uncertainty quantification are highly anticipated.

 \begin{figure*}[t]
	\centering
	\includegraphics[width=0.8\textwidth]{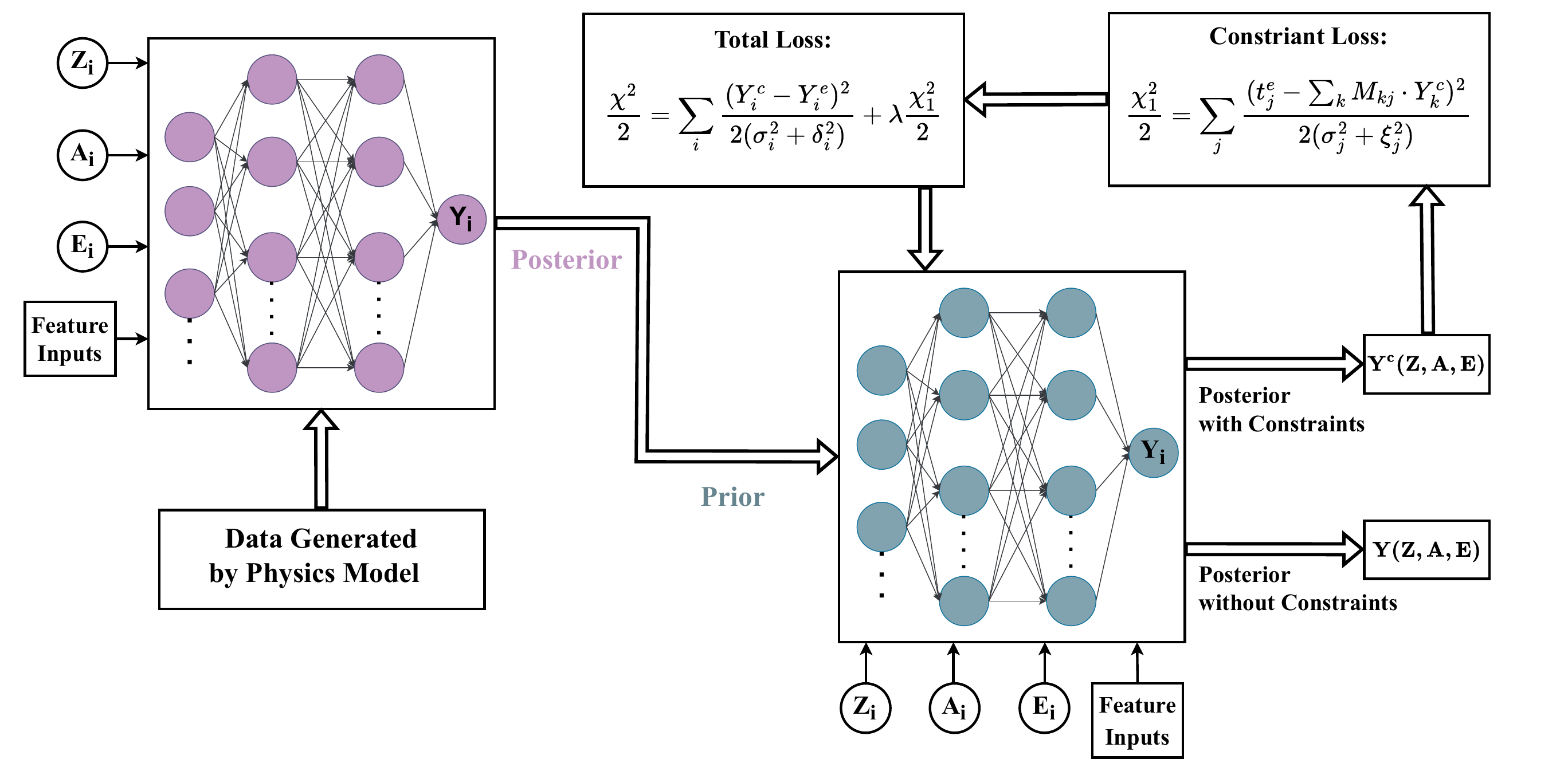}	
	\caption{ Illustration of the physics model informed Bayesian machine learning for evaluations of independent fission yields, with and without physics constraints.
    The physics model generated data are used to train the informed priors for evaluation of measured data.
    The heterogenous cumulative yields  are employed to provide additional constraints on energy dependencies via a conversion matrix. See details in the text. }
	\label{Fig1}
\end{figure*}

In this Letter, we aim to infer the energy dependence and uncertainties of independent fission yields
with physics informed Bayesian machine learning. The independent yields
have fewer measured data than cumulative yields, but it is important for fuel cycle operations and burnup calculations in nuclear energy productions
and also for understandings of fission processes~\cite{nucleardataneeds}. Previously we
have performed inferences on cumulative yields using Bayesian neural networks (BNN)~\cite{wangza1,wangza2}.
The obtained results can capture the energy dependence to some extent but not satisfactory
about fine structures in yield distributions.
In addition to BNN, there have been several machine learning approaches for evaluations of fission yields~\cite{densitymix1,densitymix2,yansw,song,schnabel}.
It will be more difficult to evaluate independent yields, regarding the scarcity of data.
To integrate machine learning and physics knowledge, the GEF model~\cite{GEF} with energy dependence
has been employed to provide informative priors. The GEF model provides general descriptions of fission observables
with parameterized fission mechanism such as shell effects, level densities,
odd-even effects in yields, and multi-chance fission~\cite{GEF}. Based on physics informed priors,
the truly Bayesian inference after learning the existing data is achieved.
Furthermore, the cumulative yields, connected with independent yields via a conversion matrix,  are employed as a physics constraint to 
 supplement energy dependence into independent yields.

 \emph{Methods.}---
The BNN framework~\cite{bnn} has wide applications in nuclear physics since 2016~\cite{utama},
which is now improved by GPU-accelerated training.
The basic inputs are the fragment mass number $A$, the fragment charge number $Z$, and the neutron incident energy $E$.
The output is the yield $Y_i(A_i, Z_i, E_i)$ of each fragment.
The physics informed Bayesian inference, including physics constraints,  is illustrated in Fig.\ref{Fig1}.
Firstly,  GEF model is used with default parameters to generate the energy dependent neutron-induced independent yields $D_{\rm phys}(A_i, Z_i, E_i, Y_i)$ of
$^{235}$U.
The posterior $P(w_1|D_{\rm phys})$ after learning $D_{\rm phys}$ with 10$^5$ steps
is provided as the prior $P(w_2)$ for evaluating experimental measured data $D_{\rm expt}$, which can be written as:
\begin{equation}
\begin{array}{l}
P(w_2)=P(w_1|D_{\rm phys}) = \displaystyle \frac{P(D_{\rm phys}|w_1)P(w_1)}{\int P(D_{\rm phys}|{w_1})P(w_1) dw_1} \vspace{8pt} \\
P(w_2|D_{\rm expt}) =  \displaystyle  \frac{P(D_{\rm expt}|w_2)P(w_2)}{\int P(D_{\rm expt}|{w_2})P(w_2)d w_2}  \vspace{8pt} \\
P(Y_n) =  \displaystyle  \int f_{\rm net}(X_n, w_2) P(w_2|D_{\rm expt}) d w_2
\end{array}
\label{eq1}
\end{equation}
After learning the measured data, the updated posterior $P(w_2|D_{\rm expt})$  and the network function $f_{\rm net}(X_n, w_2)$ are used to infer fission yields $Y_n$ with interpolation or extrapolation inputs $X_n$.
In the BNN, the Markov chain Monte Carlo is used to sample the distribution of network weights~\cite{bnn}.
The uncertainties of the inferences are given as the confidence interval (CI) at the $95\%$ level.
The network function is determined by the network architecture.
The neural network has two hidden layers and each has 22 neurons, in which
the activation function adopts the tanh function.
Based on these steps, the physics model is integrated into the  prior of BNN, which is similar to the transfer learning~\cite{transfer}.
To our knowledge, BNN with physics informed priors has not been applied in nuclear physics before.
More sophisticated models such as the Langevin approach~\cite{langevin} and TD-GCM~\cite{tdgcm} can also be incorporated by learning model-generated data.
It has to be noted that the overfitting is no longer an issue for deeper networks by using physics informed priors and sufficient model-generated data.

 \begin{figure}[ht]
	\centering
	\includegraphics[width=0.5\textwidth]{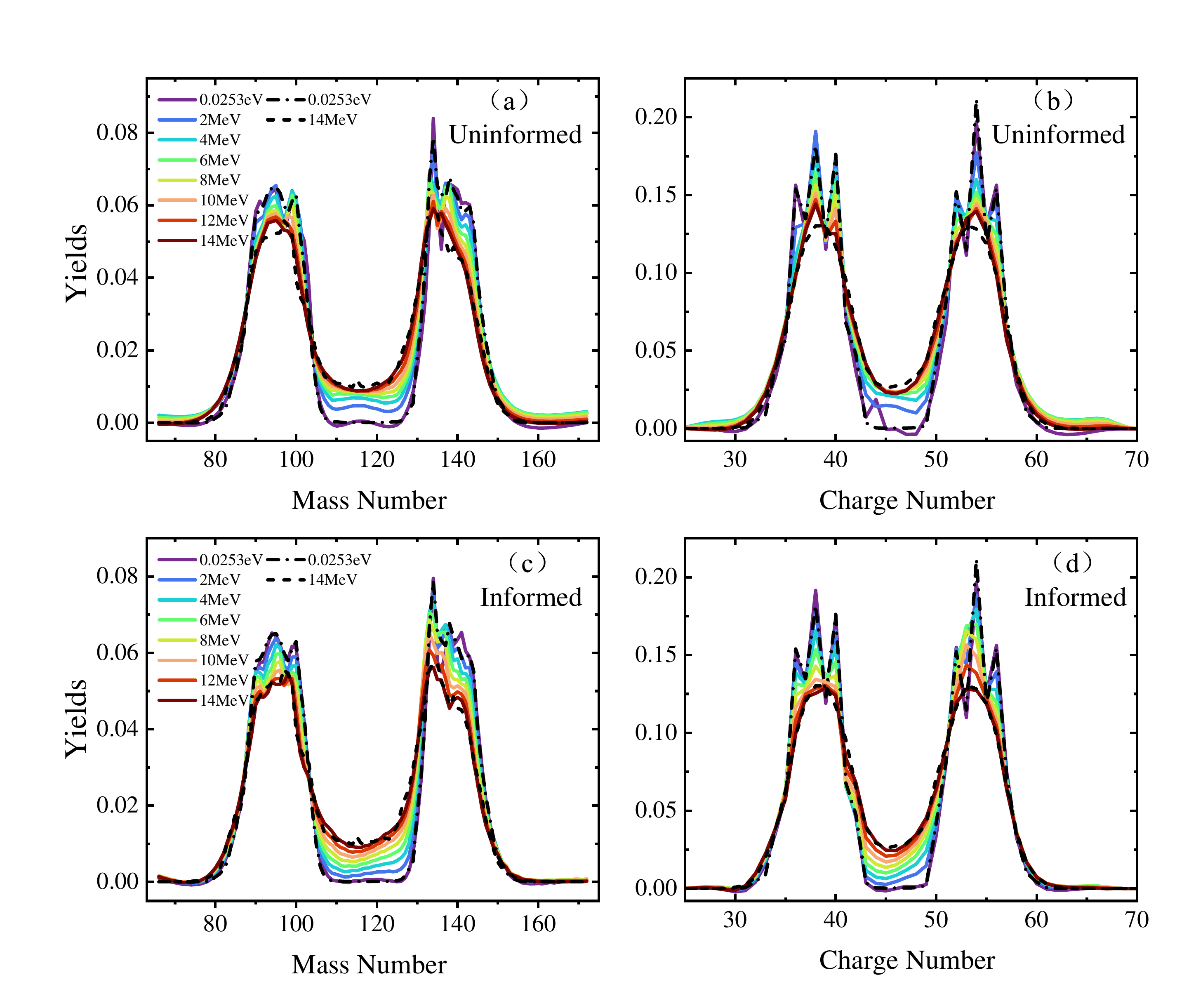}	
	\caption{
The inferences of neutron induced fission yields of $^{235}$U at different incident energies.
 (a)  mass yields with uninformed learning; (b) charge yields with uninformed learning;
 (c)  mass yields with informed learning; (d)  charge yields with informed learning.
 The evaluated JENDL data (black lines) at 0.0253 eV and 14 MeV are also shown for comparison. }
	\label{Fig2}
\end{figure}

To overcome the scarcity of independent yields, it is natural to take into account cumulative yields that have more data points
to improve the energy dependence.
Since independent and cumulative yields share common variables, 
i.e., the nuclear mass number
and neutron incident energies, the heterogenous  data fusion has been applied~\cite{wangza3}.
Presently these two kinds of yields are directly connected by the conversion matrix~\cite{matrix}, which is mainly determined
by $\beta$-decays.
The cost function with this constraint is written as
\begin{equation}
\frac{\chi^2 }{2}= \sum_i \frac{(Y_i^{\rm c}-Y_i^{\rm e})^2}{2(\sigma^2+\delta_i^2)} +
 \sum_j \lambda \frac{(t_j^{\rm e}-\sum_k M_{kj}Y_k^{\rm c})^2}{2(\sigma^2+\xi_j^2)}
\end{equation}
where the measured independent and cumulative yields are $Y_i^{\rm e}$ and $t_j^{\rm e}$, respectively,
and corresponding experimental uncertainties are $\delta_i^2$ and $\xi_j^2$.
The independent yields predicted by the network are denoted as  $Y_i^{\rm c}$.
The global noise scale is $\sigma^2$, which is determined by the learning process.
The experimental noise scale can be adjusted regarding the weights of specific points.
The likelihood function $P(D|w)$ in Eq.(\ref{eq1}) is given by ${\rm expt(-\chi^2/2)}$.

For the neutron induced fission of $^{235}$U, there are about 587 measured data points of independent yields and 1521 measured points of cumulative yields in
the Exfor library~\cite{exfor}.
Since our objection is to infer the energy dependence, the evaluated independent yields from JENDL-5 at thermal energy, 0.5 MeV and 14 MeV are also included.
We also employed 9330 data points generated by the GEF model with neutron energies of 0.0, 0.6, 2.0, 4.0, 6.0, 8.0, 10.0, 12.0, 14.0 MeV for training the priors.
In addition to  physics informed priors and physics constraints,
inputs with physics features can also be helpful, which has been discussed in our earlier work~\cite{qiao1,qiaonst}.

 \emph{Results.}---
The applications need fission yields in terms of both fragment mass
and fragment charge. Thus the machine learning of  two-dimensional yield distributions in terms of ($A$, $Z$)
 is performed, which is more difficult compared to the learning of one-dimensional yields in earlier studies~\cite{wangza1}.
In particular, the charge yields have evident odd-even staggering effects~\cite{expt-oddeven}, as explained by the double projection on particle numbers of fragments~\cite{schunck1}.
The one-dimensional mass or charge yields are obtained by summing the two-dimensional yields.

Fig.\ref{Fig2} presents the inferences of neutron induced independent fission yields of $^{235}$U.
Fig.\ref{Fig2}(a,b) shows the results from basic BNN evaluations of mass yields
and charge yields, referred as uninformed learning, based on measured data and evaluated data, with neutron incident energies from 0 to 14 MeV.
Fig.\ref{Fig2}(c,d) shows the corresponding results of BNN with the GEF model informed priors, referred as informed learning.
Generally the informed learning can better reproduce the evaluated data.
The normalization deviation from  uninformed learning is about 5.3\%, while it is about 0.22\% for informed learning.
It can be seen that the energy dependencies between two approaches are distinctly different.
It is known that symmetric fission channel increases with increasing excitation energies due to the decreasing asymmetric fission barriers~\cite{sheikh}.
For uninformed learning, the yields of symmetric channel increases quickly at low
energies but then increases slowly towards 14 MeV. 
This is not reasonable as the symmetric channel should increase exponentially and slowly at low energies~\cite{energyd}.

 \begin{figure}[b]
	\centering
	\includegraphics[width=0.5\textwidth]{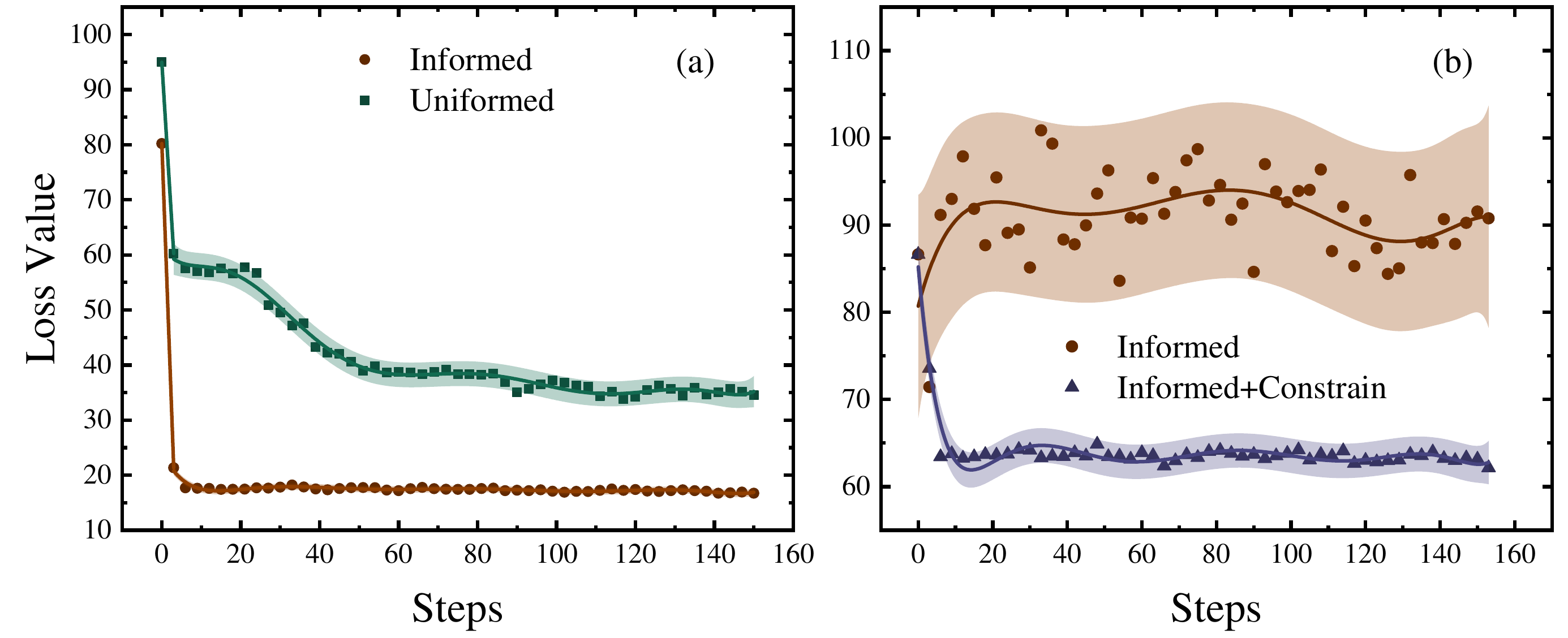}	
	\caption{(a) The loss values with and without informed priors as a function of learning steps;
(b) the loss values of informed learning  with and without constraints with respect to the target cumulative yields.
The loss values are displayed in every 3 steps for smoothing results.
The shadow region denotes the distribution of data points associated with a polynomial fitting curve. The loss values are given in
units of particles per 100 fissions (PC/FIS).
 }
	\label{Fig3}
\end{figure}

 \begin{figure}[t]
	\centering
	\includegraphics[width=0.5\textwidth]{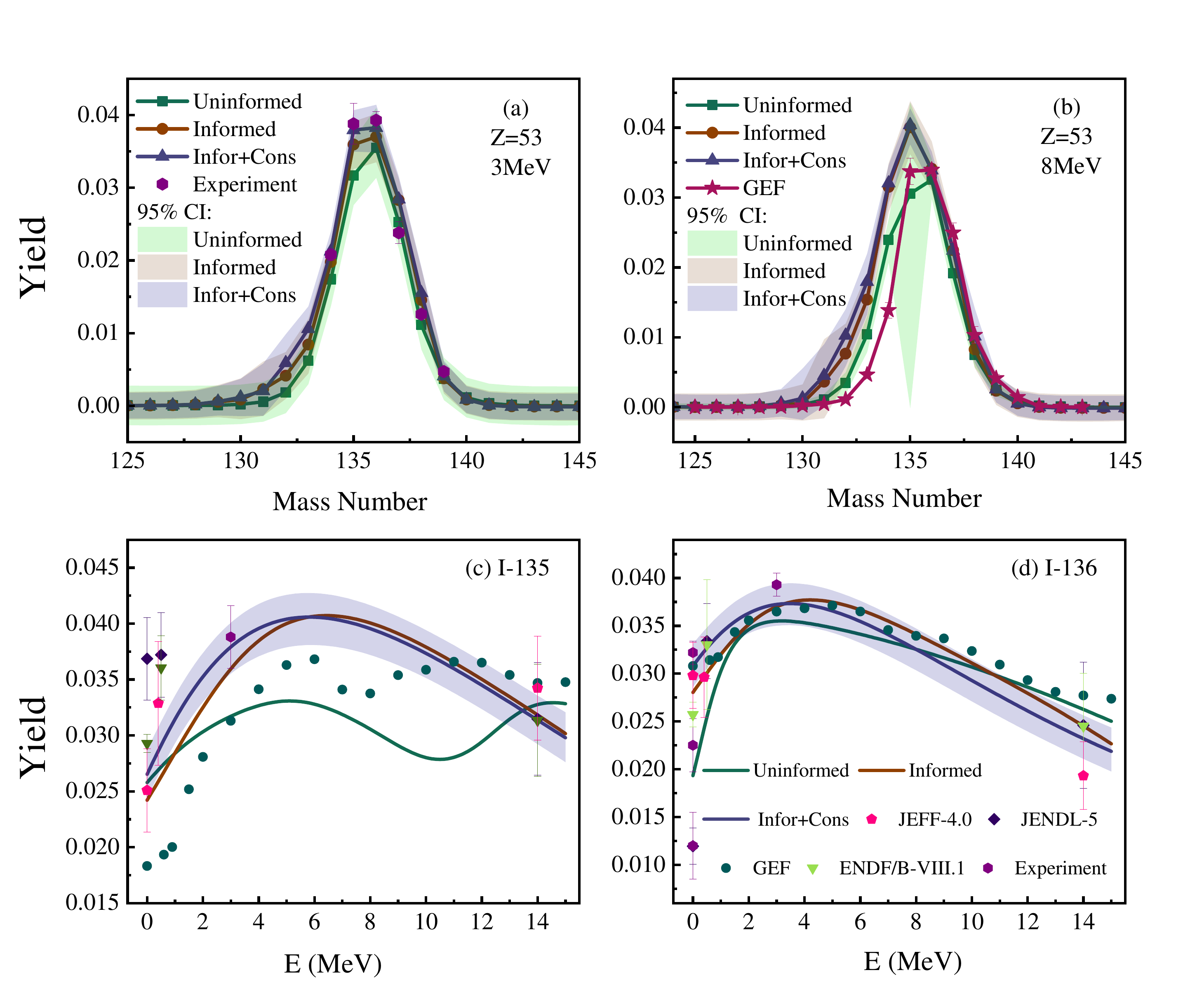}	
	\caption{ (a, b) The independent yields of $Z$=53 isotopes at 3 and 8 MeV, respectively, in which
   different approaches, such as the uninformed learning, informed learning, the informed learning with constraints,
   the GEF results, are employed.
   (c, d) The energy dependent yields of $^{135, 136}$I with different approaches. The experiment data and different evaluation data
   are also shown.  }
	\label{Fig4}
\end{figure}

For the distributions of mass yields, the peaks at $A$=95 and 100 shown at thermal energy almost disappear at 14 MeV, while
the peak at $A$=134 still exists at 14 MeV although it has been reduced.
For uninformed learning, both the peak at $A$=134 and the dip at $A$=136 are overestimated, compared to evaluated data.
For the distributions of charge yields, the evolution of the odd-even staggering is also different within two approaches.
It is noticeable that the dip at $Z$=53 disappears quickly in informed learning and the peaks at $Z$=52 and 54 are merged into one peak
towards 14 MeV gradually.
It is still challenging for theoretical models to describe these fine structures.
With informed learning, the odd-even staggering disappears at 10 MeV for light fragments and at 6 MeV for heavy fragments.
The odd-even staggering disappears later with uninformed learning. 
The pairing correlations should be much reduced at temperatures around 0.7 MeV~\cite{goodman}, indicating the odd-even staggering will disappear at low excitation energies.
Since GEF has taken into account multi-chance fission and empirical energy dependence~\cite{GEF},
we see that the evolution pattern and fine structures can be reasonably interpolated by the GEF informed learning.

To illustrate the  training performances with and without physics priors, Fig.\ref{Fig3} shows
the convergence of different approaches.
In Fig.\ref{Fig3}(a), the loss values with respect to the JENDL evaluated data are shown as a function
of learning steps. With the GEF informed prior, we see the convergence can be obtained quickly.
In the uninformed learning, the convergence is improving slowly with large uncertainties.
It is understandable that the transfer learning can drastically speed up the training of similar data~\cite{transfer}.
In Fig.\ref{Fig3}(b), the loss values with respect to the target values of measured cumulative yields with $Z$=52 to 54 are displayed.
The loss values without constraints are widely distributed and insensitive to learning steps,  and are about 50\% higher than that with constrained learning.
This demonstrated that the constraint can quickly take effect associated with small uncertainties.
The loss values with constraints cannot be further reduced, which is mainly because of the discrepant experimental data of cumulative yields.

Fig.\ref{Fig4} displays the detailed  energy dependencies of fission yields around $Z$=53 fragments.
Fig.\ref{Fig4} (a,b) shows the fission yields of $Z$=53 isotopes at 3 and 8 MeV,
using different approaches. We see that the inferences without physics priors
are generally lower and  associated with large uncertainties, in particular at 8 MeV, in contrast to informed learning.
 The experimental yields at 3 MeV~\cite{3mev}  can be
well described by informed learning.
Around 8 MeV, the GEF results are also lower than informed learning.
The yield-energy relations of $^{135,136}$I(Z=53) are shown in Fig.\ref{Fig4} (c,d).
For $^{135}$I at the peak of $Z$=53 isotopic yields, the inferences with physics priors
shows a non-monotonic energy dependencies and result in higher yields than that from uninformed learning.
It seems that the informed learning is much affected by the experimental data at 3.0 MeV~\cite{3mev}.
This explains the large charge yields at $Z$=53 as shown in Fig.\ref{Fig2}(d).
For comparison, the non-monotonic energy dependencies in $^{136}$I are similar with all approaches. 
More results are provided in the supplemental material~\cite{supplement}.
The inferences with and without physics constraints are close, indicating
independent yields are mostly compatible with cumulative yields.
However, the constraints indeed play a role since 
the constrained loss values of $Z$=53 isotopes with and without constraints are 0.17 and 0.57, respectively.
Nevertheless, 
there are considerable discrepancies between JENDL-5~\cite{jendl}, ENDF/B-VIII.1~\cite{endf} and JEFF-4.0~\cite{jeff} evaluations at the thermal energy and 14 MeV,
and further efforts are needed. In this respect, the combined use of surrogate reactions, inverse kinematics, and magnetic spectrometers
has made significant progress in measuring complete isotopic frisson yields~\cite{romas}.

 \emph{Conclusions.}---
This work tackled the bottleneck problem in purely data-driven machine learning of nuclear data due to the absence
of physics knowledge and the scarcity of data, by leveraging the physics-model informed Bayesian machine learning approach.
We focus on the evaluation and inference of energy dependent independent fission yields, for which
measured data are extremely sparse, while both theories and experiments have large uncertainties.
 The GEF model generated data is used for the pre-training
of physics informed priors. Actually more sophisticated models can also be incorporated into the priors.
Compared to  earlier approaches using physics feature inputs and physics constraints, physics informed priors can incorporate comprehensive and consistent physics knowledge
containing conservation laws and quantum effects.

The physics informed learning enables detailed studies of fine structures and energy dependencies in distributions of
fission yields.
The supply of high quality fission yields is important for  nuclear energy productions using new types of
fuels such as the thorium fuel and mixed fuels.
The inferences with and without informed priors have distinctly different energy dependencies,
 which is attributed to the non-monotonic energy dependence obtained from
informed learning, demonstrating that informed learning is essential for
reliable evaluations.
The physics constraints from cumulative fission yields are also included to provide augmented energy dependencies.

This work focuses on the evaluation of fission yields, but evidently the
physics informed Bayesian machine learning can improve the inferences of nuclear structure and reaction observables when only small
or imperfect datasets are available, such
as reaction cross sections and the equation of state of nuclear matter.
We expect that machine learning in nuclear physics will be greatly boosted in forthcoming years by incorporating
existing nuclear theories and models.


\par

\acknowledgments
We thank useful discussions with F.R. Xu and Z.G. Ge.
 This work was supported by  the
 National Key R$\&$D Program of China (Grant 2023YFE0101500, No.2023YFA1606403),
  the National Natural Science Foundation of China under Grants No.12475118, 12335007,12275081.

\end{document}